\documentclass[english,twocolumn,aps,prd,longbibliography,crop,tikz,superscriptaddress]{revtex4-1}
\usepackage[T1]{fontenc}
\usepackage{braket}
\usepackage{amsmath}
\usepackage{xcolor}
\usepackage{algorithm}
\usepackage{algorithmic}
\usepackage{amssymb}
\usepackage{graphicx}
\usepackage{babel}
\usepackage{qcircuit}
\usepackage{verbatim} 
\usepackage{lipsum}
\usepackage{blindtext}
\usepackage{upgreek}
\usepackage{soul}
\usepackage{caption}

\usepackage{fancyhdr}

\fancypagestyle{fancyplain}
\DeclareMathAlphabet{\mathcal}{OMS}{cmsy}{m}{n}
\SetMathAlphabet{\mathcal}{bold}{OMS}{cmsy}{b}{n}

\AtBeginDocument{%
  }

\begin{document}

\title{Quantum circuit knitting using a weak-coupling approximation }

\author{John P. T. Stenger}
\thanks{Corresponding Author: john.p.stenger2.civ@nrl.navy.mil}
\affiliation{Chemistry Division, U.S. Naval Research Laboratory, Washington, DC, United States}

\author{Daniel Gunlycke}
\affiliation{Chemistry Division, U.S. Naval Research Laboratory, Washington, DC, United States}

\author{Nikos Chrisochoides}
\affiliation{Depart. of Computer Science and Physics, Old Dominion University,Norfolk, VA, United States
}

\begin{abstract}
  We present a method for performing distributed quantum computing with controlled approximations for circuits that are weakly coupled via a single qubit.  Exact distributed quantum computing requires exponential  classical information to reconstruct the quantum process.  However, we show how the classical cost is reduced to polynomial if the quantum procedure can be partitioned between a qubit that is weakly coupled to the other qubits.  We demonstrate our method for a layered circuit based on the circuits used for the quantum approximate optimization algorithm.  
\end{abstract}

\maketitle
\thispagestyle{fancyplain}

\section{Introduction}

Quantum computers are predicted to provide an exponential reduction in compute time compared to classical computers for certain tasks~\cite{Feynman1982}.  Although modern quantum computers only realize this advantage in specialized cases~\cite{Youngseok2026,Arute2020,Abanin2025}, quantum computers are continuing to advance, as larger and more error resistant quantum computers are developed~\cite{Preskill2018,Arute2019,Graham2019,Bruzewicz2019,kjaergaard2020}.  However, there will always be some bound on the number of qubits available in a given quantum computer.  Thus,  Distributed Quantum Computing (DQC) is a critical pathway for scaling quantum algorithms beyond the limits of monolithic quantum processing units (QPUs), even for fault tolerant quantum computers. By partitioning large quantum circuits into smaller sub-circuits, DQC enables the execution of complex algorithms on near-term hardware. There are many approaches to DQC such as circuit knitting~\cite{Ayral2020,Tang2021,Perlin2021,wei2022,Basu2024}, probabilistic DQC~\cite{Mitarai2021,Piveteau2024,Brenner2025,Harrow2025,Johnson2026}, qubit reduction techniques~\cite{Bravyi2016,stenger2022b}, entanglement foraging~\cite{Eddins2022}, tensor networks~\cite{Bravyi2016,Peng2020,Yuan2021,wei2022}, and physical considerations~\cite{Kreula2016,rubin2016,yamazaki2018,kawashima2021}, not all of which are mutually exclusive.  We focus on circuit knitting using a particular physical consideration.  Recent literature classifies circuit knitting into two primary categories: \textit{gate cutting}~\cite{Piveteau2024,Schmitt2025} and \textit{wire cutting}~\cite{brenner2023optimal,Brenner2025}. Gate cutting decomposes non-local gates into a linear combination of local operations often implemented via quasiprobability decomposition. Conversely, wire cutting involves measuring a qubit at the cut location and preparing a corresponding state on the receiving QPU. We focus on the wire cutting paradigm.  In particular we will implement a version of the CutQC method~\cite{Tang2021}.  

A major road block for circuit knitting, and DQC in general, is that it requires an exponential amount of classical information to reconstruct a general quantum circuit.
In some cases, it may be worth performing the exponentially hard classical knitting procedure.  In other cases, there are acceptable approximations that allow for sub-exponential knitting procedures.   We show how approximations can be managed when two quantum processes are weakly coupled.  Specifically when a quantum procedure is represented on two sets of qubits that are connected by only a single qubit and the coupling to that qubit is weak.  This can happen, for example, in the quantum approximate optimization algorithm~\cite{Farhi2014,Hadfield2017,Zhou2020,Zou2025} when the underlining utility function involves two sets of binary operators that are nearly independent, or in Hamiltonian-based quantum simulation~\cite{Wecker2015,Ho2019,Wiersema2020} for which the Hamiltonian can be represented in two weakly coupled Hilbert spaces.  Such subsystem separations tend to naturally occur in physical situations where two systems are nearly independent.  For example, two molecules that are weakly coupled, or the Vehicle Routing Problem (VRP) when two depots are far apart.  Another example is cavity-mediated interconnects, which allow for tunable coupling strengths where the interaction between qubit modules can be treated perturbatively \cite{tomes2026accurate,sierra2024bath}.  Weak coupling is already being used to justify partitioning in the VRP~\cite{maciejunes2025vrp} and in image processing~\cite{billias2025qhed}.

In this paper, we provide a method for approximating the CutQC procedure~\cite{Ayral2020,Perlin2021,Tang2021,wei2022,Basu2024} in the presence of weak coupling.  Although other approximation methods have been proposed based on tomography~\cite{harada2025}, entanglement foraging~\cite{Eddins2022}, tensor networks~\cite{Bravyi2016,Peng2020, Yuan2021,wei2022}, and quasiprobability decomposition~\cite{Harrow2025},  our method uses a different approximation technique that may be more viable for certain physically relevant problems.  Our technique involves a novel string-index notation that helps to organize the relevant sub-system processes based on the order of magnitude of their probabilities.  We show that for a number of cuts $N_x$ and a small coupling strength $\lambda<1/N_X$, we can reach a precision bounded by $\epsilon_n = (N_x\lambda)^{(n+1)/3}$ using a number of classical processes $N_C = \mathcal O (N_x^{n})$ with a sampling overhead of $\gamma = \mathcal O (N_x^{3n/2})$.   

\begin{figure*}[t]
  \centering
  \includegraphics[width=\linewidth]{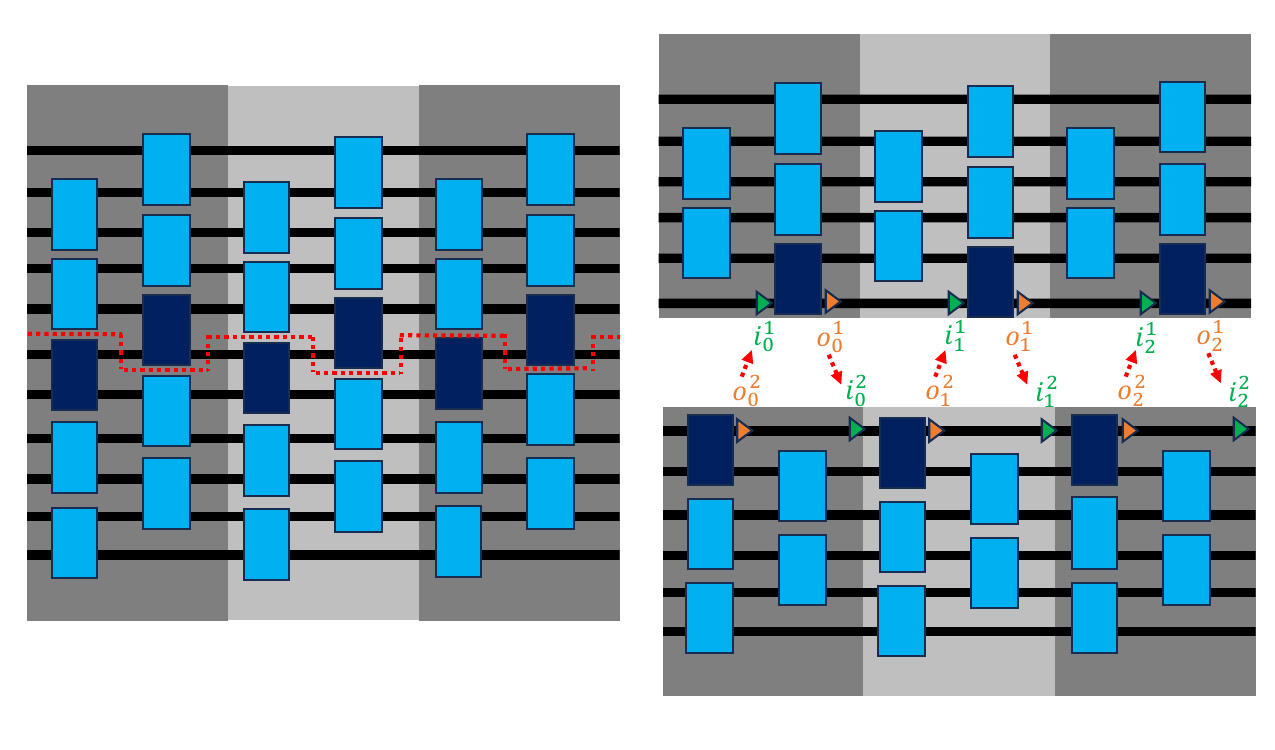}
  \caption{A quantum circuit cut along a single qubit line.    Left: full circuit.  Right: resulting subcircuits.  The solid lines represent qubits and the rectangles represent two qubit operators that may be composed of multiple one and two qubit gates.  Operators connecting to the cut qubit are drawn darker than the others representing that they are weak.  The shaded areas represent layers of the circuit.  The dashed line shows the division of the circuit.  Each time the dashed line crosses a qubit, a cut is made.  Green triangles represent input operations, which include qubit resets and single qubit gates.  Orange triangles represent output operations, which include measurement and single qubit gates.  Each input and output operation has an associated symbol that represents an instruction for the operation.  The dashed arrows indicate that the output of each circuit informs the input of the other circuit.   }
  \label{F1}
\end{figure*}

\section{Method}
\subsection{Circuit Knitting using String-Index Notation}

Let us define a set of qubits $\mathcal{Q}$ in a quantum register initialized to  
\begin{equation}
    \ket{\Phi_0} = \bigotimes_{q\in \mathcal{Q}} \ket{0}_q,
\end{equation}
where $\ket{0}_q$ represents a basis state of qubit $q$.  We prepare an ensemble of quantum registers all initialized to $\ket{\Phi_0}$.  We refer to each register as a sample and can think of each sample as occurring on the same hardware but at different times.  The initial ensemble of samples is described by the density matrix 
\begin{equation}
    \hat \rho = \ket{\Phi_0}\!\bra{\Phi_0}.
\end{equation}
A quantum circuit is applied to each sample.  The circuit can be described as an ordered list of operations $\mathcal C = (\ldots , G_b, G_a)$ where the operations are applied to the register from right to left and each operation updates the ensemble $G_i: \hat \rho \mapsto \hat \rho'$. Each operation in the circuit is either a quantum gate defined by a unitary operator, a reset operation, or a measurement operation.   We also maintain a classical register in order to record the results of the measurements.  

We divide the quantum register into three sets 
\begin{equation}
    \mathcal{Q} = \mathcal{Q}_1\cup\{q_x\}\cup \mathcal{Q}_2,
\end{equation}  
such that the sets are  mutually disjoint, $q_x \notin \mathcal{Q}_1$, $q_x \notin \mathcal{Q}_2$ and $\mathcal{Q}_1\cap \mathcal{Q}_2 = \emptyset$.   We restrict $\mathcal C$ so that all gates are of the form $\hat G_i = \hat g_{1x} \otimes \hat{\mathbb{I}}_2$ or $\hat G_j = \hat {\mathbb{I}}_1 \otimes \hat g_{x2}$ where $\hat g_{1x}$ acts on qubits in $\mathcal Q_1\cup{q_x}$, $\hat g_{x2}$ act on qubits in ${q_x}\cup\mathcal{Q}_2$, and $\hat{\mathbb{I}}_{1}$, $\hat{\mathbb{I}}_2$ are identity operators acting on qubits in $\mathcal{Q}_{1}$, $\mathcal{Q}_2$ respectively.  In this way, no gate couples $\mathcal{Q}_1$ and $\mathcal{Q}_2$ directly, but some gates do couple $\mathcal{Q}_1$ or $\mathcal{Q}_2$ to $q_x$.    The circuit is cut $N_x$ times along $q_x$ so that the circuit is divided into two sub-circuits $\mathcal{C}_1$ and $\mathcal{C}_2$, as shown in Fig.~\ref{F1}. Note that qubit $q_x$ can be coupled to any qubit in $\mathcal{Q}_1$ or $\mathcal{Q}_2$, not necessarily only adjacent qubits as shown in the figure. 

We label qubits in the subcircuits with the same label as the corresponding qubits in the initial circuit.  In this way, $q_x$ refers to a qubit line in both subcircuits.  To uniquely identify a qubit, the circuit must also be specified.  For each cut there is an output operation acting on $q_x$ in one circuit and an input operation acting on $q_x$ in the other circuit.

The output operations are performed before each cut.  They involve a measurement of $q_x$ in a certain basis. For cut $l \in \{0,\ldots, N_x-1\}$ and circuit $c \in \{1,2\}$, a symbol $o_l^c \in \{X,Y,Z,\mathbb{I}\}$ determines the basis for the measurement, where $X,Y,Z$ refer to Pauli operators and $\mathbb{I}$ refers to the identity operator.  Let $ M_{o_l^c}$ be the output operation given the symbol $o_l^c$.  The output operations are
\begin{equation}
\begin{split} 
    & M_{X}: \hat \rho \mapsto \sum_{\nu} \hat \Pi^X_{\nu} \hat \rho \hat \Pi^X_{\nu}, \\
    & M_{Y}: \hat \rho \mapsto \sum_{\nu} \hat \Pi^Y_{\nu} \hat \rho \hat \Pi^Y_{\nu} ,\\
    & M_{Z}: \hat \rho \mapsto \sum_{\nu} \hat \Pi^Z_{\nu} \hat \rho \hat \Pi^Z_{\nu} ,\\
    & M_{\mathbb{I}}: \hat \rho \mapsto \sum_{\nu} \hat \Pi^Z_{\nu} \hat \rho \hat \Pi^Z_{\nu},
\end{split}
\end{equation}
where $\hat \Pi^X_{\nu}$, $\hat \Pi^Y_{\nu}$, $\hat \Pi^Z_{\nu}$ project to the $\nu \in \{0,1\}$ eigenstate of the Pauli operators $\hat X_{q_x}$,$\hat Y_{q_x}$,$\hat Z_{q_x}$ acting on $q_x$.  Specifically $\hat \Pi^P_{\nu} = \ket{P_{\nu}}\!\bra{P_\nu}$ where 
\begin{equation}
    \begin{split}
        &\ket{Z_0} \equiv \ket{0}_{q_x},\\
        &\ket{Z_1} \equiv \ket{1}_{q_x} = \hat X_{q_x} \ket{0}_{q_x}, \\
        &\ket{X_{0,1}} \equiv \ket{\pm}_{q_x} = \big(\ket{0}_{q_x} \pm \ket{1}_{q_x}\big)/\sqrt{2},\\
        & \ket{Y_{0,1}} \equiv \ket{\pm i}_{q_x} = \big(\ket{0}_{q_x} \pm i \ket{1}_{q_x}\big)/\sqrt{2}
    \end{split}
\end{equation}
 The measurement operations also return an expectation value of the corresponding Pauli operator or identity operator to the classical register.

The input operations are performed after each cut.  They involve a reset of $q_x$ in a particular basis.  There is a symbol $i_l^c \in \{0,1,+,i\}$ for cut $l$ and circuit $c$ that determines the basis.  Let $R_{i_l^c}$ be the reset operation given the symbol $i_l
^c$. The result of the corresponding reset operations is
\begin{equation}
    \begin{split}
        &R_0 : \hat \rho \mapsto  \sum_{\nu} \bra{\nu}\hat \rho \ket{\nu}_{q_x}  \ket{0}\!\bra{0}_{q_x}
        \\
        &R_1 : \hat \rho \mapsto  \sum_{\nu} \bra{\nu}\hat \rho \ket{\nu}_{q_x}  \ket{1}\!\bra{1}_{q_x}
        \\
        &R_+ : \hat \rho \mapsto  \sum_{\nu} \bra{\nu}\hat \rho \ket{\nu}_{q_x}  \ket{+}\!\bra{+}_{q_x}
        \\
        &R_i : \hat \rho \mapsto  \sum_{\nu} \bra{\nu}\hat \rho \ket{\nu}_{q_x}  \ket{i}\!\bra{i}_{q_x}
    \end{split}
\end{equation}
where $\nu \in \{0,1\}$.  

\begin{figure*}[t]
  \centering
  \includegraphics[width=\linewidth]{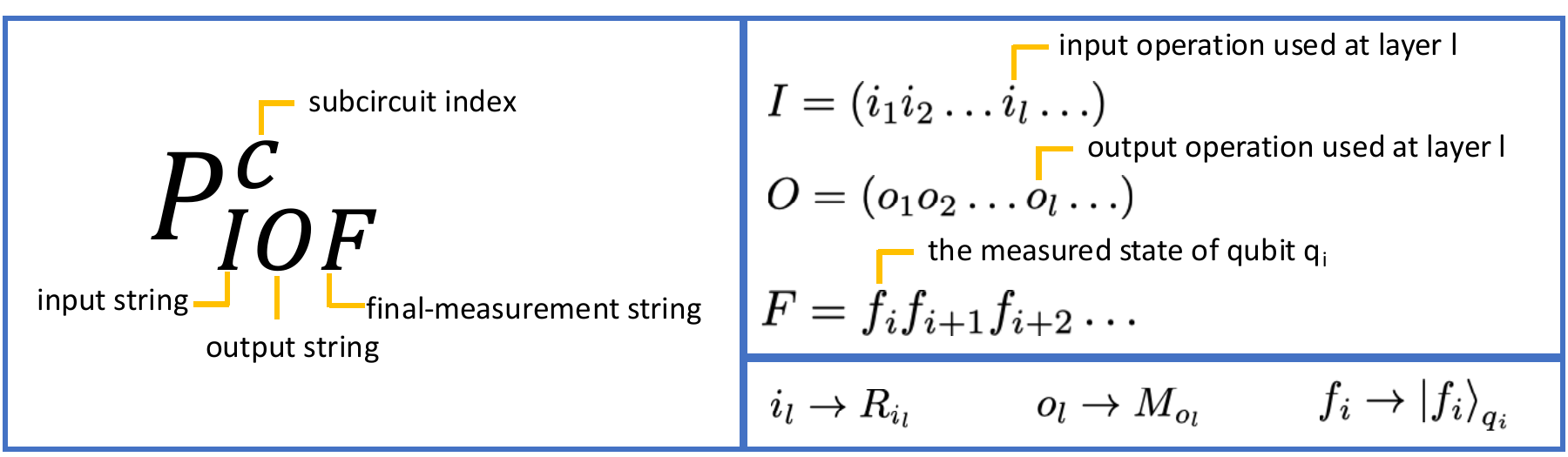}
  \caption{ String-index notation.  The probabilities are indexed by input strings, output strings, and final-measurement strings.  The input strings label the input operations, the output strings label the output operations, and the final-measurement strings label the state that is returned from the final measurement.  }
  \label{FN2}
\end{figure*}

Every unique combination of input and output operations defines a circuit and each circuit has an associated probability.  We want to formulate a method for choosing which circuits has the highest probabilities so that we can choose to run only the most relevant circuits.  To formulate such a method, we first devise a notation that helps us to track the characteristics of each circuit.  We use the output and input symbols to form output strings $O^c = (o^c_1 o^c_2 o^c_3 \ldots)$ and input strings  $I^c = (i^c_1 i^c_2 i^c_3 \ldots)$.  Let us also define the final-measurement-result string $F = f_i f_{i+1}f_{i+2}\ldots$ where $f_j \in \{0,1\}$ represents the state of the qubit $q_j$ after the circuit has been executed and measured in the Z-basis.  Let $F=F^1 F^2$ where $F^c$ contains the final results for circuit $c$.  Note that the final results for $q_x$ might be included in $F^1$ or $F^2$ depending on whether $N_x$ is even or odd.   We define the result
$ P^{c}_{I O F} $ as the probability that circuit $c$ will return a final measurement $F$ weighted by the expectation value of every measurement operator defined by $O$ given that the resets defined by $I$ are used.  Figure~\ref{FN2} provides an overview of the notation used to label the probabilities.  We drop the $c$ label on the strings when it is implied by the label on the probability.   Throughout, we use the convention for $P^c_{ABC}$ that the first string $A$ is the input string, the second $B$ is the output string, and the third $C$ is the final-measurement-result string.  

We want to knit the probabilities $ P^{c}_{I O F} $ together so that we obtain the full-circuit probability.  However, knitting requires that we have output in the same basis as the input. Let us label objects with symbols from $\{0,1,+,i\}$ as $\alpha$-type objects and those from $\{X,Y,Z,\mathbb{I}\}$ as $\sigma$-type objects.  We cannot perform $\alpha$-type measurements directly, but we can use the $\sigma$-type probabilities to determine $\alpha$-type probabilities algebraically.  Let us define a transformation matrix $\Gamma$ such that
\begin{equation}
\label{eq7}
    P^c_{I O_\alpha F} = \sum_{O_\sigma} \Gamma_{O_\alpha}^{O_\sigma} P^c_{I O_\sigma F},
\end{equation}
where $O_\sigma$ are $\sigma$-type output strings and $O_{\alpha}$ are $\alpha$-type output strings.  Correspondingly, we define $\sigma_l$ to be the $\sigma$-type output symbols, and $\alpha_l$ to be the $\alpha$-type output symbols.  The matrix $\Gamma$ applies the necessary algebra to transform $\sigma$-type probabilities into $\alpha$-type probabilities.  

Let us order the strings as $\vec{O}_{\alpha} = \bigotimes_l (0,1,+,i)$ and $\vec{O}_{\sigma} = \bigotimes_l (\mathbb{I},Z,X,Y)$, where we are generalizing the kronecker product for vectors of strings, e.g. $(a,b)\bigotimes(c,d) = (aa,ad,bc,bd)$.   Using this ordering, we can derive the inverse transformation matrix
\begin{equation}
    \Gamma^{-1} = \bigotimes_l 
    \begin{pmatrix}
    \braket{0|\hat{\mathbb{I}} |0} & \braket{1|\hat{\mathbb{I}} |1} & \braket{+|\hat{\mathbb{I}} |+} & \braket{i|\hat{\mathbb{I}} |i} \\
    \braket{0|\hat Z |0} & \braket{1|\hat Z |1} & \braket{+|\hat Z |+} & \braket{i|\hat Z |i} \\
    \braket{0|\hat X |0} & \braket{1|\hat X |1} & \braket{+|\hat X |+} & \braket{i|\hat X |i} \\
    \braket{0|\hat Y |0} & \braket{1|\hat Y |1} & \braket{+|\hat Y |+} & \braket{i|\hat Y |i} \\
    \end{pmatrix},
\end{equation}
where we have dropped the $q_x$ labels on the operators and the states for readability.  This gives us
\begin{equation}
    \Gamma^{-1} = \bigotimes_l 
    \begin{pmatrix}
    1 & 1 & 1 & 1 \\
    1 & -1 & 0 & 0 \\
    0 & 0 & 1 & 0 \\
    0 & 0 & 0 & 1
    \end{pmatrix},
    \label{Gm1}
\end{equation}
which has the solution
\begin{equation}
    \Gamma = \frac{1}{2^{N_x}}\bigotimes_l 
    \begin{pmatrix}
    1  & 1  & -1 & -1 \\
    1  & -1 & -1 & -1 \\
    0 & 0 & 2 & 0 \\
    0 & 0 & 0 & 2
    \end{pmatrix}.
    \label{G}
\end{equation}
This solution holds as long as we include every string for both types of output.  However, in the following, we remove certain strings, in which case the inverse is solved numerically.

Using the $\alpha$-type symbols, we can knit the probabilities of the subcircuits to calculate the full-circuit probabilities   
\begin{equation}
    P_{F} = \sum_{\alpha_1,\alpha_2,\ldots}\sum_{\alpha_1',\alpha_2',\ldots} P^1_{(\alpha_1 \alpha_2 \ldots)(\alpha_1' \alpha_2' \ldots)F^1} P^2_{(\alpha_1' \alpha_2' \ldots)(\alpha_1 \alpha_2 \ldots)F^2},
    \label{pknit}
\end{equation}
where both $\alpha_i, \alpha_i' \in \{0,1,+,i\}$.

\subsection{Weak-Coupling Approximation}
\label{WCA}
Let us assume that $q_x$ is weakly coupled to the rest of the system by a general set of gates, as exemplified by Fig~\ref{F1}.  Any quantum circuit can be decomposed into a set of two qubit gates and single qubit gates~\cite{Lloyd1995}.  Furthermore, any two qubit gate can be decomposed into a set of single qubit gates and a two qubit gate of the form
\begin{equation}
    \hat G_{q_i,q_j} = \eta \hat{\mathbb{I}} + \lambda \hat O_i 
    \hat  O_j,
    \label{eq9}
\end{equation}
where $\hat O_i$ is an operator that acts only on $q_i$, $|\hat O_i| = 1$, $\eta$ is a complex coefficient, and $\lambda$ is a positive real number.  Appendix~\ref{ABb} shows how such a gate may originate from a physical model.  When $\lambda$ is small, $\hat G_{q_i,q_j}$ weakly couples $q_i$ to $q_j$.    

We assume the circuit is divided into layers such that in each layer, we apply both $\hat G_{q_x,q_i}$ and $\hat G_{q_j,q_x}$ where $q_i \in \mathcal{Q}_1$ and $q_j \in \mathcal{Q}_2$.  Each time $\hat  G_{q_x,q_i}$ or $\hat G_{q_j,q_x}$ is applied, we must cut $q_x$ twice.   

\begin{figure*}[t]
  \centering
  \includegraphics[width=0.5\linewidth]{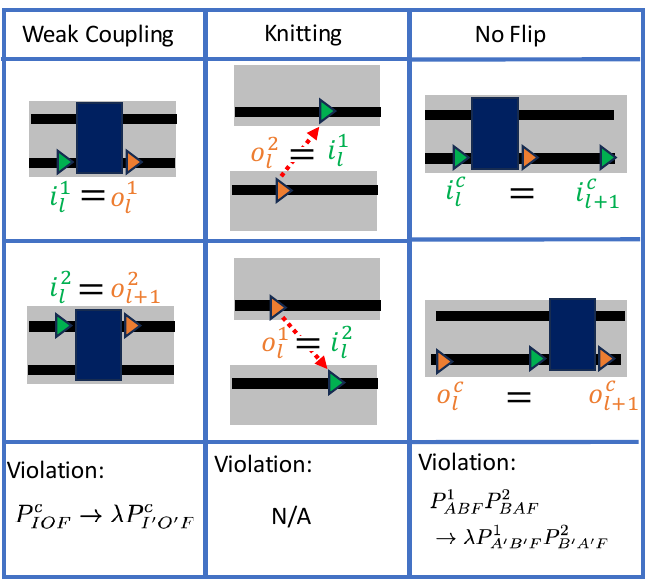}
  \caption{ Snippets of the circuit that are relevant for the three rules.  Each column is labeled by a rule and shows examples of where the rule applies in the circuit.  The bottom of each column states what happens when the rule is violated.} 
  \label{FN1}
\end{figure*}

The probability that $q_x$ changes state when either $\hat G_{q_x,q_i}$ or $\hat G_{q_j,q_x}$ is applied is $\mathcal O(\lambda)$, as shown in Appendix~\ref{ABb}. We use this fact to derive rules for the symbols $o_l^1$, $i_l^1$,$o_l^2$, $i_l^2$.  Figure~\ref{FN1} depicts a set of three rules.  As a direct consequence of the weak coupling in the circuit, we have the weak-coupling rule
\begin{equation}
\label{r1}
\begin{split}
    &o^1_{l} = i^1_{l},  \\
    &o^2_{l+1} = i^2_{l}, ~ o^2_0 = 0,
\end{split}
\end{equation}
which determines how the symbols change across gates.
 When the weak-coupling rule is obeyed, the gates in a given layer do not alter the state of the cut qubit.  Every time this rule is broken, the probability is reduced to approximately $\lambda$ times the original value.  When $\lambda$ is small, the probability rapidly decays each time the weak-coupling rule is broken.  
 
 We want to use the weak coupling rule to formulate a rule that applies to the input and output strings of each circuit independently.  To do this, we use Equation~\ref{pknit}, which implies the knitting rule
\begin{equation}
\label{r2}
\begin{split}
    & i^2_{l} = o^{1}_{l} ,\\
    & i^1_{l} = o^{2}_{l} ,
\end{split}
\end{equation}
which determines how the symbols behave across cuts.
The knitting rule is never broken.
The two rules Eq.~\eqref{r1} and \eqref{r2} together imply the no-flip rule  
\begin{equation}
\label{r3}
    \begin{split}
        &i^c_{l+1} = i^c_l, ~ i^c_0 = 0 , \\
        &o^c_{l+1} = o^c_l, ~ o^c_0 = 0,
    \end{split}
\end{equation}
which determines how symbols change in a given string.
Breaking the no flip rule reduces the probability by a factor of $\lambda$.
This means that input and output strings are exponentially suppressed by the number of times the symbols change as read from left to right.  For example, $P^1_{(00++)(0+++)F^1}P^2_{(0+++)(00++)F^2} = \mathcal O(\lambda)$ and $P^1_{(0+11)(00+1)F^1}P^2_{(00+1)(0+11)F^2} = \mathcal O(\lambda^2)$.  We refer to a change of symbol in a string as a flip. 

We can choose an approximation level $n$ and drop probabilities with magnitude $\mathcal O(\lambda^{n})$ or less.  To do this, we keep a set $\mathcal S^n_{\alpha}$ of $\alpha$-type strings for which the symbols flip no more than $n$ times.  We also keep a reduced set of $\sigma$-type strings $\mathcal S_{\sigma}^n$.  The number of $\sigma$-type strings that must be kept is the same as the number of $\alpha$-type strings, but one must be careful to keep a set of $\sigma$-type strings so that $\Gamma$ preserves normalization.  A procedure for choosing $\sigma$-type strings is given in Appendix~\ref{AB}.  Briefly, for every added $\alpha$-string $O_{\alpha}$, we select a unique $\sigma$-string $O_{\sigma}$, based on the location of flips in $O_{\alpha}$.  This ensures that the rows of $\Gamma$ remain linearly independent.  We refer to this selection procedure as Positional-Symbolic Correspondence (PSC).

For each approximation level $n$, the number of strings 
\begin{equation}
\label{eq:NS}
    N_C = \sum_{m\leq n} 3^{m} {N_x \choose m} = \mathcal O(N_x^{n})
\end{equation} 
that must be kept increases polynomially with $N_x$.  
We define the approximate full-circuit probability  
\begin{equation}
\label{eq14}
    P^{n}_{F} = \sum_{A \in \mathcal S^{n}_{\alpha}}\sum_{A' \in \mathcal S^{n}_{\alpha}} P^{n;1}_{AA'F_1} P^{n;2}_{A'AF_2},
\end{equation}
where 
\begin{equation}
P^{n;c}_{IO_{\alpha}F} = \sum_{O_\sigma \in \mathcal S_{\sigma}^n} \Gamma_{O_{\alpha}}^{n;O_\sigma} P^c_{IO_{\sigma}F},
\end{equation}
and $\Gamma_{O_{\alpha}}^{n;O_\sigma}$ is the result of inverting Eq.~\eqref{Gm1} with rows and columns removed so that the input and output strings are confined to $\mathcal S^n_{\alpha}$ and $\mathcal S^n_{\sigma}$.  Because we only drop small $\alpha$-type probabilities, we know that the approximate full-circuit probability will be close to the exact full-circuit probability.  We find the full-circuit error
\begin{equation}
    \epsilon_n \equiv |P_{F}-P^{n}_{F}| = \mathcal O\big(\bar{\lambda}^{n+1}),
    \label{bound}
\end{equation}
where $\bar{\lambda}$ depends on both $N_x$ and $\lambda$.  If we had access to the exact $\alpha$-type probabilities then $\bar \lambda = N_x \lambda$, however, we have to approximate the $\alpha$-type probabilities from the truncated set of $\sigma$-type probabilities.  We find $\bar{\lambda} \leq (N_x\lambda)^{1/3}$ using PSC for the most general case, see Appendix~\ref{ette} for details.  In practice, $\bar{\lambda}$ can be smaller than this general bound based on the specifics of the circuit.

\subsection{Local Rotations of the Cut Qubit}

Above, we assumed that $q_x$ is only acted on by two-qubit gates.  In general, $q_x$ may also be acted on by single qubit gates.  Let $\hat{H}_{l}$ be a single qubit unitary operator acting on $q_x$ during layer $l$.   We keep track of the total single qubit operation
\begin{equation}
    \bar H_l = \prod_{k \leq l} \hat{H}_{l-k}.
\end{equation}
We apply the inverse of this total operator as part of the output measurement, then re-apply the operator during input.  The measurement operations become 
\begin{equation}
\begin{split} 
    & M_{X}: \hat \rho \mapsto \sum_{\nu} \hat \Pi^X_{\nu} \bar H_{l} \hat \rho \bar H_{l}^{\dagger} \hat \Pi^X_{\nu} ,\\
    & M_{Y}: \hat \rho \mapsto \sum_{\nu} \hat \Pi^Y_{\nu} \bar H_{l} \hat \rho \bar H_{l}^{\dagger} \hat \Pi^Y_{\nu} ,\\
    & M_{Z}: \hat \rho \mapsto \sum_{\nu} \hat \Pi^Z_{\nu} \bar H_{l} \hat \rho \bar H_{l}^{\dagger} \hat \Pi^Z_{\nu} ,\\
    & M_{\mathbb{I}}: \hat \rho \mapsto \sum_{\nu} \hat \Pi^Z_{\nu} \bar H_{l} \hat \rho \bar H_{l}^{\dagger} \hat \Pi^Z_{\nu},
\end{split}
\end{equation}
and the reset operators become
\begin{equation}
    \begin{split}
        &R_0 : \hat \rho \mapsto \sum_{\nu} \bra{\nu}\hat \rho \ket{\nu}_{q_x}  \bar H_{l}\ket{0}\!\bra{0}_{q_x}\bar H_{l}^{\dagger}
        \\
        &R_1 : \hat \rho \mapsto \sum_{\nu} \bra{\nu}\hat \rho \ket{\nu}_{q_x}  \bar H_{l}\ket{1}\!\bra{1}_{q_x}\bar H_{l}^{\dagger}
        \\
        &R_+ : \hat \rho \mapsto \sum_{\nu} \bra{\nu}\hat \rho \ket{\nu}_{q_x}  \bar H_{l}\ket{+}\!\bra{+}_{q_x}\bar H_{l}^{\dagger}
        \\
        &R_i : \hat \rho \mapsto \sum_{\nu} \bra{\nu}\hat \rho \ket{\nu}_{q_x}  \bar H_{l}\ket{i}\!\bra{i}_{q_x}\bar H_{l}^{\dagger}
    \end{split}
\end{equation}
This can be understood as working in a frame that is rotating with the cut qubit.  

\section{Computational Complexity Analysis}
\label{sec:complexity}

In this section, we analyze the end-to-end computational and physical sampling complexity of the weak-coupling wire-knitting protocol. The wire-cutting  procedure consists of 
three modular phases: Classical Selection, Quantum Execution, and Classical Reconstruction.  We prove that both classical processing and quantum hardware sampling scale polynomially in the weak-coupling regime.
Algorithm~\ref{alg:weak_knitting} formalizes the wire-cutting our protocol.

\begin{algorithm}[H]
\caption{Weak-Coupling Wire Knitting with Truncated Classical Reconstruction}
\label{alg:weak_knitting}
\begin{algorithmic}[1]
\REQUIRE Quantum circuit $\mathcal{C}$, interaction cut line $q_x$ with $N_x$ cuts, coupling parameter $\lambda$, truncation order $n$, shot allocation $S$.
\ENSURE Reconstructed full-circuit distribution $\hat{P}_F^n$.

 \textbf{// PHASE I: Classical String Selection }
\STATE Generate restricted set of $\alpha$-type strings $\mathcal{S}_\alpha^n$ with flip count $m \le n$. \COMMENT{Size: $N_C = \sum_{m=0}^n 3^m \binom{N_x}{m} = \mathcal{O}(N_x^n)$}
\STATE Construct corresponding $\sigma$-type Pauli strings $\mathcal{S}_\sigma^n$ via Positional-Symbolic Correspondence (PSC).
\STATE Compute truncated reconstruction matrix: \\ $\Gamma^n = \left(\left.\Gamma^{-1}\right|_{\mathcal{S}_\sigma^n \times \mathcal{S}_\alpha^n}\right)^{-1}$.

\textbf{// PHASE II: Quantum Hardware Execution }
\FORALL{selected configuration pairs $(I \in \mathcal{S}_\alpha^n, O_\sigma \in \mathcal{S}_\sigma^n)$}
    \STATE Execute Subcircuit $\mathcal{C}_1(I, O_\sigma)$ on QPU for $S$ shots; record empirical frequencies $\tilde{P}_{I O_\sigma F^1}^1$.
    \STATE Execute Subcircuit $\mathcal{C}_2(O_\sigma, I)$ on QPU for $S$ shots; record empirical frequencies $\tilde{P}_{O_\sigma I F^2}^2$.
\ENDFOR \COMMENT{Total physical budget: $S_{\text{total}} = 2 N_C S$}

 \textbf{// PHASE III: Classical Reconstruction }
\FORALL{$c \in \{1, 2\}$}
    \STATE Transform measurement bases:  \\ $\hat{P}_{I O_\alpha F^c}^{n; c} \leftarrow \sum_{O_\sigma \in \mathcal{S}_\sigma^n} (\Gamma^n)_{O_\alpha}^{O_\sigma} \tilde{P}_{I O_\sigma F^c}^c$.
\ENDFOR
\STATE Stitch subcircuit distributions: \\ $\hat{P}_F^n \leftarrow \sum_{A \in \mathcal{S}_\alpha^n} \sum_{A' \in \mathcal{S}_\alpha^n} \hat{P}_{A A' F^1}^{n; 1} \hat{P}_{A' A F^2}^{n; 2}$. \COMMENT{Classical runtime: $\mathcal{O}(N_C^2) = \mathcal{O}(N_x^{2n})$}
\RETURN $\hat{P}_F^n$
\end{algorithmic}
\end{algorithm}
There are two sources of error in the algorithm, the shot noise $\epsilon_{S_\text{total}} = \text{Var}(P^n_F)$ and the approximation error $\epsilon_n = |P_F - P^n_F|$.  The scaling of the algorithm depends on three factors, the number of strings $N_C$ generated in Phase I, the total number of samples $S_{\text{total}}$ collected in Phase II, and the reconstruction time in Phase III.  
In the presence of weak coupling 
$$ \lambda < \frac{1}{N_x} $$
the entire algorithm scales polynomially with the number of cuts
\begin{enumerate}
    \item \textbf{Number of Circuits:} The number of retained classical subcircuit strings is strictly polynomial: $N_C = \sum_{m=0}^n 3^m \binom{N_x}{m} = \mathcal{O}(N_x^n)$.
    \item \textbf{Approximation Error:} The algorithmic truncation error over the full distribution is bounded by $\epsilon_n \le \mathcal{O}[(N_x \lambda)^{(n+1)/3}]$.
    \item \textbf{Classical Post-Processing:} The classical reconstruction runtime scales as $\mathcal{O}(N_C^2) = \mathcal{O}(N_x^{2n})$.
    \item \textbf{Sample Variance:} The sampling overhead is polynomial: $\gamma = \mathcal O( N_x^{3n/2})$.
    \item \textbf{Sample Size:} The physical sample budget required to achieve additive statistical precision $\epsilon_{S_{\text{total}}}$ scales as $S_{\mathrm{total}} = \mathcal{O}[2N_x^{3n} / \epsilon_{S_{\text{total}}}^2]$.

\end{enumerate}
The number of circuits (1) is addressed in Sec~\ref{WCA} and found to be $N_C = \mathcal O(N_x^n)$.  It depends on the level of approximation $n$.  For weak coupling, we can choose a small value for $n$ as the approximation error (2) $\epsilon_n \leq \mathcal O[(N_x \lambda)^{(n+1)/3}]$, which is also addressed in Sec.~\ref{WCA}.  The most resource demanding step in the classical post processing (3) is the double sum over the number of circuits, which scales as $N_C^2 = \mathcal O(N_x^{2n})$.  What remains to be addressed is the sample variance (4) and the total number of samples required (5).

\begin{figure*}[t]
  \centering
  \includegraphics[width=\linewidth]{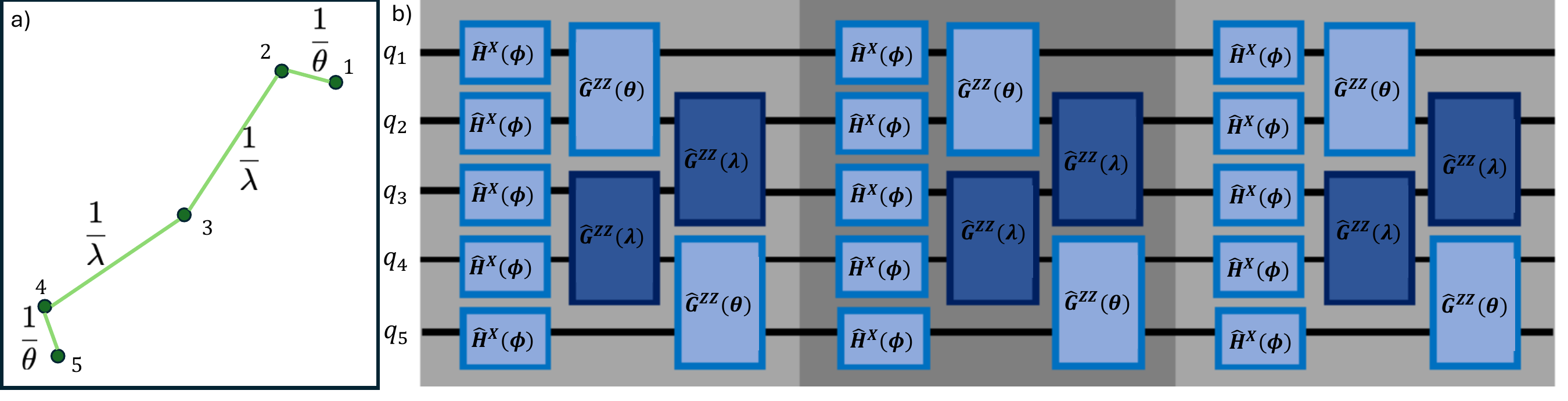}
  \caption{ (a) the graph used in the demonstration.  The nodes are numbed 1-5 and the lengths of the edges are labeled $1/\lambda$ and $1/\theta$.  Nodes 1,2 are weakly coupled to nodes 4,5 via node 3.   (b) depiction of the quantum circuit used in our demonstration.  Lines represent qubits and boxes represent quantum gates.  The qubits are labeled according to their corresponding node in (a).  Each shaded region represents one layer of the circuit.  In each layer, there is a set of single qubit $H^X(\phi)$ gates and two qubit $G^{ZZ}(\theta)$ gates.  We set $\phi = 0.1$ and $\theta = 0.5$.  The coupling $\lambda$ is varied.     }
  \label{F2}
\end{figure*}

The sample variance of a single $\sigma$-type probability is 
\begin{equation}
    \epsilon_S^2 \equiv \text{Var}(P^c_{IO_{\sigma}F}) = \frac{1}{S}
\end{equation}
where $S = S_{\text{total}}/N_C$ is the number of samples per circuit execution.  Because the transformation matrix $\Gamma^n$ preserves normalization, we expect the variance of the $\alpha$-type probabilities to be approximately the same as the variance for the $\sigma$-type probabilities
\begin{equation}
    \text{Var}(P^c_{IO_{\sigma}F}) = w^2 \epsilon_S^2,
\end{equation}
where $w = \mathcal O(1)$.
See Appendix~\ref{NS} for more information regarding the propagation of errors through the transformation matrix.  The final variance can be found from Eq.~\ref{eq:NS} using Bienaym{\'e}'s formula~\cite{feller1968introduction} and Goodman's identity for the variance of independent products~\cite{goodman1960variance,casella2002statistical}
\begin{equation}
    \epsilon_{S_{\text{total}}}^2 = \mathcal O( N^2_S \epsilon_S^2) = \mathcal O\Big(\frac{ N_x^{3n}}{S_{\text{total}}} \Big).
\end{equation}
Thus, we see that the sampling overhead is $\gamma \equiv \epsilon_{S_{\text{total}}}\sqrt{S_{\text{total}}} = \mathcal O( w N_x^{3n/2} )$.  We see that we need only a polynomial increase in the number of samples as compared to a single circuit execution to achieve the same variance.

\section{Demonstration}

We demonstrate the weak-coupling approximation on a simulated 5-qubit quantum register.  Figure~\ref{F2}a shows a 5-node graph with edges of length $1/\lambda$ and $1/\theta$ with $\lambda \ll \theta$.  A typical optimization one may perform is to minimize the edge weights.  In this case we can use the utility function operator
\begin{equation}
    \hat F = \hat Z_1 \hat Z_2 + \frac{\lambda}{\theta} \hat Z_2 \hat Z_3 + \frac{\lambda}{\theta} \hat Z_3 \hat Z_4 + \hat Z_4 \hat Z_5,
\end{equation}
To optimize this utility function,  one can use a layered quantum circuit such as the  QAOA circuit~\cite{Farhi2014}.  Of course, this particular example is optimized by $\braket{Z_{q_i}} = -1$ for all $q_i$, corresponding to zero edge weight.  However, it is not our goal here to optimize an interesting function but only to demonstrate the weak-coupling approximation on a minimal model.  The resulting QAOA circuit is initialized with a set of Hadamard gates acting on each qubit.  We apply the ansatz in layers
\begin{equation}
    \hat A_l = e^{i \zeta_l \sum_q \hat X_q} e^{i \xi_l \hat F}.
\end{equation}
We set $\zeta_0 = \phi - \pi/2$, $\zeta_{l\neq0} = \phi$, and $\xi_l = \theta$.  The initial $\zeta_0 = \phi - \pi/2$ value is set so that we can absorb the initialization into the first layer.  We apply 3 layers of the ansatz.  The resulting circuit is shown in Fig.~\ref{F2}.  During each layer, every qubit is acted on by a single qubit gate
\begin{equation}
    \hat H^X_q(\phi) = \cos(\phi) \hat{\mathbb{I}} - i \sin(\phi) \hat X_q,
\end{equation}
where we set $\phi = 0.1$.  Additionally, the qubit pairs (1,2) and (4,5) are acted on by
\begin{equation}
    \hat G^{ZZ}_{q_i,q_j}(\theta) = \cos(\theta) \hat{\mathbb{I}} - i \sin(\theta) \hat Z_{q_i} \hat Z_{q_j} ,
\end{equation}
where we set $\theta = 0.5$ and qubit pairs (2,3) and (3,4) are acted on by $G^{ZZ}_{q_i,q_j}(\lambda)$.  We cut along qubit $q_3$ and explore how the error changes with $\lambda$.  

\begin{figure*}[t]
  \centering
  \includegraphics[width=\linewidth]{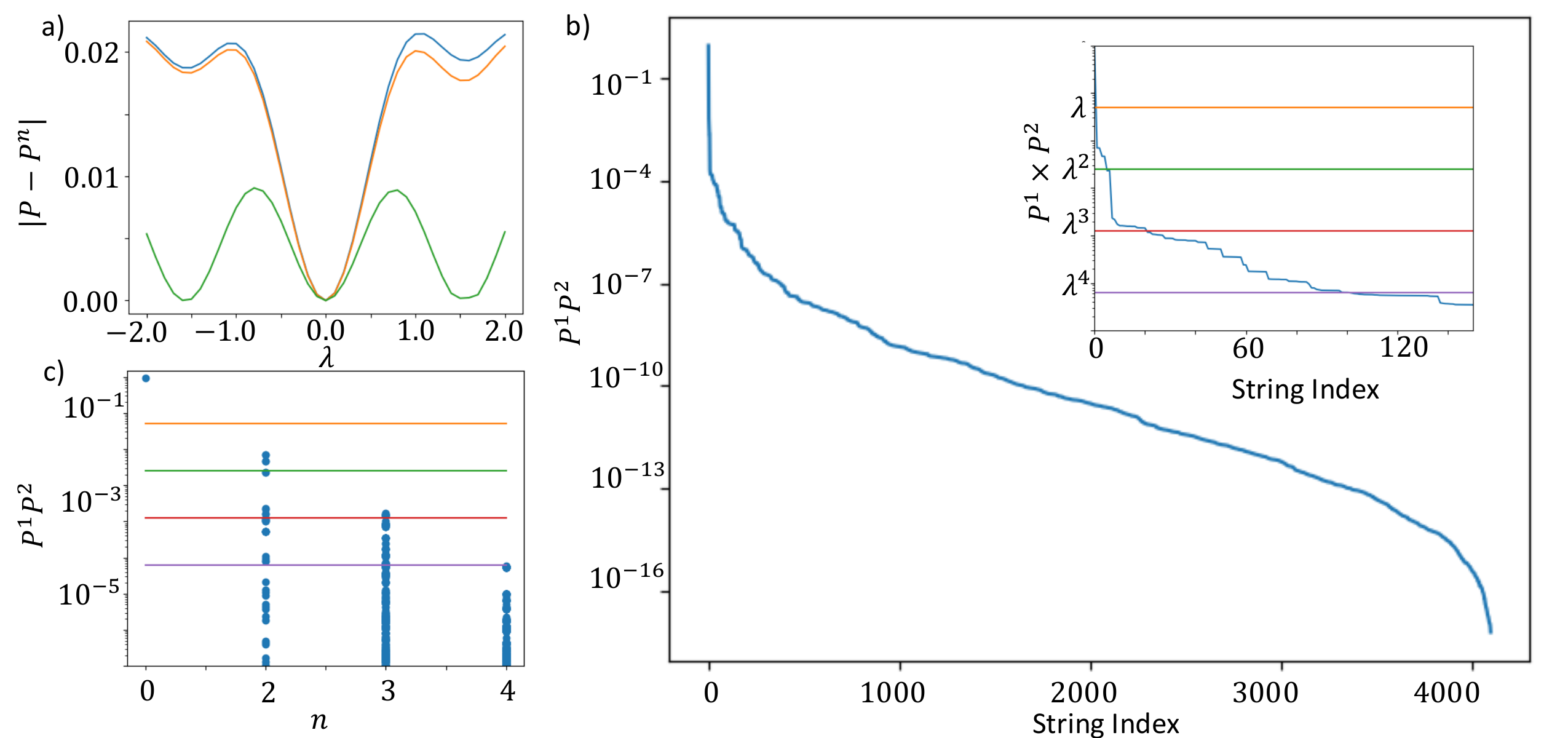}
  \caption{ Demonstration for a three layered circuit.  (a) the error in the final probability for $F={00000}$ as a function of the coupling $\lambda$ to the cut qubit $q_3$.  In order of decreasing error, the curves represent $n = 0$, $n=1$, and $n=2$.  (b) final contributed probability of each string combination.  The inset is a magnification of the same data with horizontal lines corresponding to powers of $\lambda$.  (c) final contributed probability of each string combination sorted into bins of $n$.  The horizontal lines corresponding to powers of $\lambda$. }
  \label{F3}
\end{figure*}

Figure~\ref{F3}a shows the error in the final probability of the final output $F=00000$.  The blue curve is for $n = 0$, the orange for $n = 1$, and the green for $n = 2$.  We see that the error is identically $P-P^{n} = 0$ at $\lambda = 0$ in all cases.  As $|\lambda|$ is increased, so is the error.  Away from $\lambda = 0$, the error is reduced for larger $n$.  Notice that the real error is much less than the general error bound, Eq.~\eqref{bound}, with $\bar{\lambda} = (N_x \lambda)^{1/3}$. As mentioned above, the specifics of the circuit can reduce the error.  In this case, there are two major factors that reduce the error.  Firstly, applying a single flip cannot result in the final measurement $F=00000$, and therefore there is nearly zero first-order error.  Secondly, as there are only three layers, it is not possible to pair flips as described in Appendix~\ref{ette} and so $\bar{\lambda} = \mathcal O(N_x \lambda)$ as seen by the $\lambda^2$ dependence of the second order approximation.

Figure~\ref{F3}b shows the knitted probability $P^1_{ABF}P^2_{BAF}$ for all strings $A,B$.  The strings have been index so that the probability monotonically decreases. For this plot, we set $\lambda = 0.05$.  We see that the probability exponentially decreases indicating that approximations are justified.  The inset of this figure shows a magnification of the most probable strings.  We see that there is a clustering of probabilities at powers of $\lambda^n$.  This effect is further highlighted in Fig.~\ref{F3}c where we collect the strings into bins of $n$ flips.  There is no bin for $n = 1$ as the cut qubit cannot rotate once and end in $\ket{0}_{q_3}$.  We see that the highest probabilities in each bin cluster around $\lambda^{n}$, as expected.  

\section{Discussion}
\label{sec:discussion}


The weak-coupling approximation presented here serves as a theoretical method for bypassing the exponential sampling barriers typically associated with DQC \cite{tang2025enabling}. By identifying that many practical problems possess inherent weak coupling, either through spatial locality ~\cite{billias2025qhed, maciejunes2025vrp} or clustered interactions—we render previously intractable large-scale applications accessible to small-scale quantum computing. 


Two recent applications highlight how weak coupling can be exploited in the $n=0$ limit. In combinatorial optimization, the \textit{Vehicle Routing Problem} (VRP) can be decomposed into weakly coupled clusters of cities, allowing a 156-qubit instance to be solved via parallel processing of smaller sub-circuits achieving a 96\% reduction in qubit requirements while preserving solution quality \cite{maciejunes2025vrp}. Similarly, in quantum image processing, Quantum Hadamard Edge Detection benefits from the spatial locality of pixel data. By partitioning large medical images into sub-regions, the global edge detection operator can be reconstructed from local quantum kernels with high fidelity, demonstrating utility-scale performance and effectively bypassing the noise limits of deeper, monolithic circuits\cite{billias2025qhed}.  These implementations exemplify how the theoretical weak-coupling approximation translates into massive reductions in circuit depth and gate count for real-world instances even for $n=0$.  The present work demonstrates how to go beyond $n=0$ in order to improve the approximations.

\section{Conclusion}

We present a method for dividing a quantum circuit along a weakly coupled qubit line that requires classical resources that grow polynomially with the number of cuts.  We follow the CutQC circuit knitting protocol developed in~\cite{Tang2021}, but develop our own notation that helps to illuminate the effects of weak coupling.  Our work is a specialization of the original CutQC proposal that allows one to bypass the exponential complexity of knitting for a class of physically motivated problems. General CutQC requires measurement and initialization in four different bases for each cut.  The total number of resulting circuits is $N_C = 4^{N_x}$.  We show that the probability of many of these circuits will be exponentially suppressed when the cut qubit is weakly coupled to the other qubits.  In this case, the total number of circuits that must be preserved increases polynomially with the number of cuts $N_C = \mathcal O(N_x^{n})$. 
The method is demonstrated for a layered circuit similar to that found in the quantum approximate optimization algorithm.  

 \section{Acknowledgment}
This work has been supported by the Office of Naval Research (ONR) through the U.S. Naval Research Laboratory (NRL) and for NC in part by the ONR Summer Faculty Research Program and the Richard T. Cheng Endowment at Old Dominion University. We acknowledge QC resources from IBM through a collaboration with the Air Force Research Laboratory (AFRL).

\bibliography{ref}

\appendix

\section{Weak coupling in the model}
\label{ABb}

Here we discuss how weak coupling in the model leads to the weak coupling rule defined in the main text.  Let our system be described by a utility operator $\hat F$ with an interaction term $\hat{F}_{\text{int}} $ and a coupling parameter $\lambda' < 1/N_x$.  We typically want to form unitary operators of the form 
\begin{equation}
    \hat{G}(\lambda) = \exp(i \lambda' \hat{ F }_{\text{int}} ).
\end{equation}
This type of operator is necessary in nearly all quantum algorithms, including the QAOA algorithm and for time evolution.

Assuming $\hat{ F }_{\text{int}}^2 = 1$, which can always be made true by decomposing the full utility operator into Pauli operators, we have 
\begin{equation}
    \hat{G}(\lambda) = \cos(\lambda')\hat{\mathbb{I}}  +  i\sin( \lambda' ) \hat{ F }_{\text{int}}.
\end{equation}
 This is exactly the form of Eq.~\eqref{eq9} for $\eta = \cos(\lambda')$ and $\lambda = \sin(\lambda')$. 
Physically, this expansion represents the gate as a dominant identity channel $I$ with a small perturbative interaction.

Consider the density matrix $\hat \rho$ representing the state of the system before the gate application. The evolved state $\hat \rho'$ after applying the gate $\hat{G}$ is given by
\begin{equation}
    \hat \rho' = \hat{G} \hat \rho \hat{G}^\dagger.
\end{equation}
For weak coupling $\cos(\lambda')\approx1$.  We substitute $\hat{G} \approx \hat{\mathbb{I}} - i \lambda \hat{ F }_{\text{int}} $ and its adjoint $\hat{G}^\dagger \approx \hat{\mathbb{I}} + i \lambda \hat{ F }_{\text{int}}$
\begin{equation}
    \hat \rho' \approx (\hat{\mathbb{I}} - i \lambda \hat{ F }_{\text{int}} ) \hat \rho (\hat{\mathbb{I}} + i \lambda \hat{ F }_{\text{int}} ).
\end{equation}
Expanding this product while neglecting terms proportional to $\lambda^2$
\begin{equation}
    \hat \rho' \approx \hat \rho - i \lambda \hat{ F }_{\text{int}}  \hat \rho + i \lambda \hat \rho \hat{ F }_{\text{int}}.
\end{equation}
Rearranging the terms using the commutator $[\hat{ F }_{\text{int}} , \hat \rho] = \hat{ F }_{\text{int}}  \hat \rho - \hat \rho \hat{ F }_{\text{int}} $, the evolution of the density matrix is
\begin{equation}
    \hat \rho' \approx \hat \rho - i \lambda [\hat{ F }_{\text{int}} , \hat \rho].
\end{equation}
Since the change in the density matrix $\Delta \hat \rho = \hat \rho' - \hat \rho$ is strictly proportional to $\lambda$
\begin{equation}
    \Delta \hat \rho = - i \lambda [\hat{ F }_{\text{int}} , \hat \rho],
\end{equation}
so the probability that $q_x$ changes state is also $\mathcal{O}(\lambda)$.

\section{Choosing $\sigma$-type strings}
\label{AB}

In the main text, we approximate probabilities by keeping $\alpha$-type strings with up to $n$ symbol changes.  One should keep the same number of $\sigma$-type strings, being careful that $\Gamma$ is not singular.  Enforcing that $\Gamma$ is non-singular does not uniquely define a strategy for choosing the $\sigma$-type strings. We use the following strategy, which we refer to as Positional-Symbolic Correspondence (PSC).   

At $n = 0$, the only $\alpha$-type string is $O_{\alpha} = 000\ldots$ composed of only $0$ symbols and we keep the corresponding $\sigma$-type string $O_{\sigma} = \mathbb{I}\mathbb{I}\mathbb{I}\ldots$ composed of only $\mathbb{I}$ symbols.  At higher approximation levels, for each $\alpha$-type string $O_{\alpha}$ that is added, we add a corresponding $\sigma$-type string $O_{\sigma}$ with the following rules:
\begin{enumerate}
    \item for any index $l$ such that the symbols in $O_{\alpha}$ at $l$ and $l-1$ are the same, the symbol in $O_\sigma$ at $l$ is $\mathbb{I}$ (e.g. $\ldots 11 \ldots \rightarrow \ldots ?\mathbb{I} \ldots $)
    \item at any index $l$ where $O_{\alpha}$ changes symbols such that the symbol at $l$ is $1,+,i$ the symbol at index $l$ in the corresponding string $O_{\sigma}$ is $Z,X,Y$, respectively (e.g. $ \ldots11\!+\!+\ldots \rightarrow \ldots ?\mathbb{I}X\mathbb{I} \ldots  $), 
    \item at any index $l$ where $O_{\alpha}$ changes symbols such that the symbol at $l-1$ is $1,+,i$ and the symbol at $l$ is $0$, the symbol at index $l$ in the corresponding string $O_{\sigma}$ is $Z,X,Y$, respectively (e.g. $\ldots ii00\ldots \rightarrow \ldots ?\mathbb{I}Y\mathbb{I} \ldots $),
\end{enumerate}
where $?$ indicates that the symbol depends on the details of the unspecified portion of the string.

For example,
\begin{align*}
        & O_{\alpha} = 00000 &&\rightarrow && O_{\sigma} = \mathbb{I}\mathbb{I}\mathbb{I}\mathbb{I}\mathbb{I} ,\\
        & O_{\alpha} = 00\!+\!++  &&\rightarrow && O_{\sigma} = \mathbb{I}\mathbb{I}X\mathbb{I}\mathbb{I} ,\\
        & O_{\alpha} = 011ii  &&\rightarrow && O_{\sigma} = \mathbb{I}X\mathbb{I}Y\mathbb{I} ,\\
        & O_{\alpha} = 00\!+\!+0  &&\rightarrow && O_{\sigma} = \mathbb{I}\mathbb{I}X\mathbb{I}X .\\
\end{align*}



\section{Evaluating the Total Error}
\label{ette}

We want to find an order of magnitude approximation for the total error
\begin{equation}
    \epsilon = |P_{F }-P^n_{F }| .
\end{equation}
We write this error in terms of a sum over errors 
\begin{equation}
    \epsilon = |\sum_{A,A'} \epsilon_{A A' }| ,
\end{equation}
where 
\begin{equation}
    \epsilon_{A A' } = P^1_{A A' F }P^2_{A' A F } - P^{n;1}_{A A' F }P^{n;2}_{A' A F } .
\end{equation}

Errors in the sub-circuit probability come from approximating $\Gamma$ as can be seen by considering that
\begin{equation}
    P_{I O_\alpha F }^{n;c} = \sum_{O_\sigma}(\Gamma^{n})^{O_\sigma}_{O_\alpha}P^c_{I O_\sigma F }. 
    \label{eq49}
\end{equation}
Approximating $\Gamma^n$ does not influence the  $\sigma$-type probabilities because they are measured directly from the quantum computer. Therefore, we can write the $\sigma$-type probabilities in terms of the exact $\alpha$-type probabilities
\begin{equation}
    P^c_{I O_\sigma F } = \sum_{O_\alpha}(\Gamma^{-1})^{O_\alpha}_{O_\sigma}P^c_{I O_\alpha F }.
    \label{eq50}
\end{equation}
Using Eq.~\eqref{eq50} in Eq.~\eqref{eq49} we obtain the approximate $\alpha$-type probabilities in terms of the exact $\alpha$-type probabilities
\begin{equation}
    P_{I O_\alpha F }^{n;c} = \sum_{O_\alpha'} \Upsilon_{O_\alpha}^{O_\alpha'} P^n_{I O_\alpha' F },
\end{equation}
where 
\begin{equation}
    \Upsilon_{O_\alpha}^{O_\alpha'} = \sum_{O_\sigma \in \mathcal S^n_{\sigma}}(\Gamma^{n})^{O_\sigma}_{O_\alpha}(\Gamma^{-1})^{O_\alpha'}_{O_\sigma}. 
\end{equation}
By construction, we know that if $O_{\alpha}'\in \mathcal S^n_{\alpha}$ then $\Upsilon^{O_{\alpha}'}_{O_{\alpha}} = 0$ unless $O_{\alpha}'=O_{\alpha}$ in which case  $\Upsilon^{O_{\alpha}}_{O_{\alpha}} = 1$ , therefore, we find the approximate probability in terms of the exact probability
\begin{equation}
    P_{I O_\alpha F }^{n;c} = P_{I O_\alpha F }^c + \sum_{O_\alpha' \notin \mathcal S^n_{\alpha}} \Upsilon_{O_\alpha}^{O_\alpha'} P_{I O_\alpha' F }^c.
\end{equation}
The total probability involves pairs of subcircuit probabilities

\begin{equation}
\begin{aligned}
    P_{A A' F^1 }^{n;1} P_{A' A F^2 }^{n;2}  
    &= \Big( P_{A A' F^1 }^1 + \sum_{O_\alpha \notin \mathcal{S}^n_{\alpha}} \Upsilon_{A'}^{O_\alpha} P_{A O_\alpha F^1 }^1 \Big) \\
    &\quad \times \Big( P_{A' A F^2 }^2 + \sum_{O_\alpha \notin \mathcal{S}^n_{\alpha}} \Upsilon_{A}^{O_\alpha} P_{S' O_\alpha F^2 }^2 \Big),
\end{aligned}
\end{equation}

rearranging the terms we have
\begin{equation}
\begin{split}
\epsilon_{A A' }=&P_{A A' F^1 }^1 \sum_{O_\alpha \notin \mathcal S^n_{\alpha}} \Upsilon_{A}^{O_\alpha} P_{A' O_\alpha F^2 }^2 +
P_{A' A F^2 }^2 \sum_{O_\alpha \notin \mathcal S^n_{\alpha}} \Upsilon_{A'}^{O_\alpha} P_{A O_\alpha F^1 }^1  
\\
+&\sum_{O_\alpha \notin \mathcal S^n_{\alpha}} \Upsilon_{S'}^{O_\alpha} P_{S O_\alpha F^1 }^1 \sum_{O_\alpha' \notin \mathcal S^n_{\alpha}} \Upsilon_{S}^{O_\alpha'} P_{S' O_\alpha' F^2 }^2.
\end{split}
\label{eq57}
\end{equation}
We use Eq.~\eqref{eq57} to evaluate the total error.  We do this by placing two different error bounds on Eq.~\eqref{eq57}.

Let $m$ be the number of symbol changes in $A$ and $m'$ be the number of symbol changes in $A'$.  From the no-flip rule, Eq.~\eqref{r3},  we know that $P^1_{A A' F^1 }P^2_{A' A F^2 } = \mathcal O[\lambda^{\max(m,m')}] $.  We want to use this as an error bound, however, notice that Eq.~\eqref{eq57} involves sums over $O_\alpha$.  Fortunately, because $\Upsilon^{O_\alpha'}_{O_\alpha}$ must distinguish $O_\alpha$ from all strings with $m$ or fewer flips, $\Upsilon^{O_\alpha'}_{O_\alpha} =0$ unless $O_\alpha'$ flips symbols at least at every index that $O_{\alpha}$ flips symbols.  Therefore, the bound $\mathcal O[\lambda^{\max(m,m')}] $ applies to every term in Eq.~\eqref{eq57} and, thus, $\epsilon_{A A' } = \mathcal O[\lambda^{\max(m,m')}] $.
This is the first bound and it tells us that for large $m$ or $m'$ the error is small. 

To address small $m,m'$, we use the weak-coupling rule Eq.~\eqref{r1}  and the fact that the strings $O_\alpha \notin \mathcal S^n_{\alpha}$ have at least $n+1$ symbol flips.  
We ask how many symbols must be different between $A$ and $O_\alpha \notin\mathcal S^n_\alpha$.  The minimal number of symbol differences occurs if every cut index that involves a flip in $A$ also involves the same flip in $O_\alpha$.  In this case, there are still $|n+1-m|$ flips that occur in $O_\alpha$ that are unmatched in $A$.   To minimize the symbol differences between $A$ and $O_\alpha$, we can pair these unmatched flips so that only a single symbol is different for each pair, meaning that the number of symbol differences is $|n+1-m|/2$, assuming $|n+1-m|$ is even.  Take for example, $A = 000111$ and $O_\alpha = 0\!+\!0101$.  In this example, $m = 1$ and $n+1=5$ and there are $|n+1-m|/2 = 2$ symbol differences.  Furthermore, there is no symbol $O_\alpha'$ with $n+1=5$ flips that has fewer symbol differences. Therefore, using the weak-coupling rule, we have that $P^c_{A A' F } = \mathcal{O}(\lambda^{|m-m'|/2})$  and $P^c_{A O_{\alpha}\notin \mathcal S^n_{\alpha} F } = \mathcal{O}(\lambda^{|n+1-m|/2})$.  We can replace the probabilities in Eq.~\eqref{eq57} with their order of magnitude 

\begin{equation}
\begin{aligned}
    \epsilon_{A A' } 
    &\lesssim (\lambda^{|m-m'|/2})(\lambda^{|n+1-m'|/2}) \\
    &\quad + (\lambda^{|m'-m|/2})(\lambda^{|n+1-m|/2}) \\
    &\quad + (\lambda^{|n+1-m|/2})(\lambda^{|n+1-m'|/2}).
\end{aligned}
\end{equation}

The largest values occur for $m=m'$ in which case  $\epsilon_{A A' } = \mathcal O( \lambda^{|n+1-m|/2})$.  This is our second bound on $\epsilon_{A A' }$ and it addresses the case when both $m$ and $m'$ are small. 

According to these two bounds, the largest error occurs for intermediate values of $m$ and $m'$.  Specifically, the bounds cross at $m= (n+1-m)/2$ at which point
the errors are $\mathcal O(\lambda^{(n+1)/3})$ and there are approximately $N_x^{(n+1)/3}$ such terms.  Thus, the total error is bounded 
\begin{equation}
    \epsilon = \mathcal O[(N_x\lambda)^{(n+1)/3}] .
\end{equation}

\section{Numerical stability}
\label{NS}

\subsection{Bounded Propagation of Quantum Hardware Noise and Sample Variance}
\label{remark:noise_propagation}
In addition to bounding the algorithmic truncation error, the reconstruction matrix $\Gamma$ robustly prevents exponential amplification of native quantum hardware noise or sample variance. If the raw quantum processor yields measurements with an error margin $\chi$, the final reconstructed probability distribution will only suffer a propagated error of $\omega \lesssim w\chi$, where $w$ is a small integer strictly bounded by the PSC rules. 

A persistent challenge in quantum error mitigation is that inverting coefficient matrices often amplifies hardware noise exponentially. However, our truncated reconstruction matrix avoids this because its non-zero elements are structurally constrained to small fractions ($\pm w/ N_{O_\alpha^c}$) derived from the PSC. When calculating the propagated error, summing these fractional weights over the relevant subsets cancels the large $N_{O_\alpha^c}$ denominators, capping the noise amplification by a small integer factor $w$. Thus, $\Gamma$ inherently acts as a stabilizing filter rather than a noise amplifier, ensuring the final reconstructed probabilities roughly preserve the order of magnitude of the native quantum hardware error.

To formally demonstrate this bounded propagation, consider that there is an amount of error coming from a combination of hardware errors and sample varriance so that that the measured probabilities are
\begin{equation}
    \tilde P^c_{I O_{\sigma} F } = P^c_{I O_{\sigma} F } + \chi^c_{O_{\sigma}},
\end{equation}
for some error $\chi^c_{O_{\sigma}}$.  The error propagates into the final probabilities through $\Gamma$.  We can  calculate the propagated error from Eq.~\eqref{eq7} 
\begin{equation}
   \omega^c_{O_\sigma} = \sum_{O_{\alpha}} \Gamma_{O_\sigma}^{n;O_{\alpha}} \chi^c_{O_{\alpha}}.
\end{equation}
Assuming we have kept a set of $\sigma$-type strings $S^n_{\sigma}$, we know that $\Gamma_{n;O_\sigma}^{O_{\alpha}} = 0$ unless $O_\sigma$ corresponds with $O_{\alpha}$ such that $+\mapsto X$ and $i \mapsto Y$.  Let $\mathcal S^{n}_{O_{\alpha}} \subset \mathcal S^n_{\alpha}$ be the set of strings in $S^n_{\alpha}$ that correspond as described above and let $N_{O_{\alpha}} = |S^{n}_{O_{\alpha}}|$ be the number of such strings.  Then from Eq.~\eqref{Gm1} we know the non-zero values are $\Gamma_{O_\sigma}^{O_{\alpha}} = \pm w^c_{O_\alpha}/N_{O_{\alpha}} $ where $w^c_{O_\alpha}$ is an undetermined integer.  Thus, the error is
\begin{equation}
   \chi^c_{O_\sigma} = \sum_{O_{\alpha} \in \mathcal S_{n}^{O_{\alpha}}} \frac{\pm w_{O_\alpha}}{N_{O_{\alpha}}} \chi^c_{O_{\alpha}} \leq w \chi^c,
\end{equation}
where  $|\chi^c| \geq  |\chi^c_{O_{\alpha}}|$ is the largest magnitude among the errors and $w \geq w_{O_\alpha}$ is the largest value of $w_{O_\alpha}$.  If all of the strings are kept, then $w_{O_\alpha} = 2$ for all $O_\alpha$, as seen from Eq.~\eqref{G}.  When the strings are chosen using PSC, numerical results suggest that $w_{O_\alpha} \leq 3$.  Thus, we see that $\Gamma$ roughly preserves the error due to the quantum computer.

\subsection{Preservation of Normalization}
\label{remark:prob_preservation}
While the reconstruction matrix $\Gamma$ in general is not strictly diagonally dominant, it inherently preserves the total probability of the quantum system. Specifically,  
\begin{equation}
\sum_{O_{\alpha}}\sum_{F}P^c_{I O_{\alpha} F } = 1.
\end{equation}
This is true for $\Gamma$ at any approximation level
\begin{equation}
\label{eq:preservation}
\sum_{O_{\sigma}}\sum_{O_{\alpha}}\sum_{F} \Gamma_{O_\sigma}^{n;O_\alpha} P^c_{I O_{\sigma} F } = 1.
\end{equation}
In other words, $\Gamma^n$ always preserves normalization.  This can be seen by inverting Eq.~\eqref{eq7}
\begin{equation}
     P^c_{I O_{\sigma} F } =  \sum_{O_\alpha} (\Gamma^{-1})^{O_{\alpha}}_{O_{\sigma}} P^c_{I O_{\alpha} F }.
\end{equation}
We can define a new set of probabilities $ P^{n;c}_{I O_{\alpha} F }$ such that we force the probabilities for strings outside of $\mathcal S^n_{\alpha}$ to be zero and 
\begin{equation}
     P^c_{I O_{\sigma} F } =  \sum_{O_\alpha \in S^n_{\alpha}} (\Gamma^{-1})^{O_{\alpha}}_{O_{\sigma}}  P^{n;c}_{I O_{\alpha} F } ~\text{for}~O_{\sigma} \in \mathcal S^n_{\sigma}, 
     \label{55}
\end{equation}
where $|\mathcal S^n_{\sigma}| = |S^n_{\alpha}|$.
In order for Eq.~\eqref{55} to hold,  $ P^{n;c}_{I O_{\alpha} F }$ must be normalized.  Because the probabilities outside of $\mathcal S^n_{\alpha}$ are zero, we can drop those terms in the summation 
\begin{equation}
\begin{aligned}
    P^c_{I O_{\sigma} \in S^n_{\alpha} F } 
    &= \sum_{O_\alpha} (\Gamma^{-1})^{O_{\alpha}}_{O_{\sigma}} P^{n;c}_{I O_{\alpha} F } \\
    &= \sum_{O_\alpha \in S^n_\alpha} (\Gamma^{-1})^{O_{\alpha}}_{O_{\sigma}} P^{n;c}_{I O_{\alpha} F } \\
    &= \sum_{O_\alpha \in S^n_\alpha} \big[(\Gamma^n)^{-1}\big]^{O_{\alpha}}_{O_{\sigma}} P^{n;c}_{I O_{\alpha} F }.
\end{aligned}
\end{equation}

Thus $\Gamma^n$ preserves normalization.

\subsection{Bounded Infinity-Norm}
\label{remark:infinity_norm}
The truncated matrix $\Gamma^n$ is defined over the polynomial subset $S$, forming an $N_C \times N_C$ matrix, where $N_C = \mathcal O(N_x^n)$  from Eq. \ref{eq:NS} ; with an infinity-norm bounded by:
\begin{equation}
    \|\Gamma^n\|_\infty \le N_C \|\Gamma^n\|_1. \label{eq:norm_equivalence}
\end{equation}
Since the subset size is finite—scaling only polynomially with the number of cuts $N_x$ under the weak-coupling approximation—and the $1$-norm $\|\Gamma^n\|_1$ is bounded, see Eq. \ref{eq:preservation} , it follows that the infinity-norm $\|\Gamma^n\|_\infty$ is also strictly bounded ~\cite{Golub2013}. 

To formally address the numerical stability of the classical post-processing, we must evaluate the condition number of the reconstruction matrix, defined with respect to the infinity-norm as $\kappa_\infty(\Gamma^n) = \|\Gamma^n\|_\infty \|(\Gamma^n)^{-1}\|_\infty$. 

We know the $\|\Gamma^n\|_\infty$ is bounded. 
Because $\Gamma^n$ represents a full-rank, non-singular mapping over a finite vector space, its minimum singular value is strictly bounded away from zero and thus the infinity-norm of the inverse matrix, $\|(\Gamma^n)^{-1}\|_\infty$, is finite and bounded ~\cite{Golub2013}.  Thus, the condition number is bounded:
\begin{equation}
    \kappa_\infty(\Gamma^n) \le \mathcal{O}(N_C) \cdot \|(\Gamma^n)^{-1}\|_\infty, 
    \label{eq:condition_number}
\end{equation}
since $N_C = \mathcal{O}(N_x^n)$, the condition number of the truncated reconstruction matrix scales at most polynomially with the number of cuts $N_x$. Therefore, even in the general case, i.e., without strict diagonal dominance, $\Gamma$ is well-conditioned. 

\subsection{Balancing approximation errors and sample variance}
\label{baeasv}
A critical consideration for NISQ implementation is the interplay between the approximation error, $\epsilon =|P_{F }-P^{n}_{F }|$ and the statistical error.  While higher approximation orders $n$ theoretically reduce $\epsilon$, they also increase the sampling overhead $N_C$, thereby reducing the number of samples available for repeated measurements.  Using $N_C \approx N_x^n$ and $\epsilon_n \approx (N_x \lambda)^{n/3}$ we have $N_C \approx N_x^{3\log_{N_x\lambda}(\epsilon_n)}$.  Let us use the example of $\lambda = 1/N_x^2$ so that $\epsilon_n^3 \approx 1/N_C$. We can model the total error 
\begin{equation}
    \epsilon_{\text{total}} \approx \underbrace{\frac{1}{N_C^{1/3}}}_{\text{Approx. Error}} + \underbrace{(1 - F_{\text{gate}}^{N_{\text{gate}}}) + R}_{\text{Hardware Error}} + \underbrace{\frac{\sqrt{N_C}}{\sqrt{S_{\text{total}}}}}_{\text{Statistical Error}},
\end{equation}
where $\lambda$ is the coupling strength, $F_{\text{gate}}$ is the average gate fidelity, $N_{\text{gate}}$ is the circuit depth, $R$ is the readout error, and $S_{\text{total}}$ is the total sample budget~\cite{Piveteau2024}, so that $S_{\text{total}}/N_C$ is the number of samples used for repeated measurements for each measurement string.  The gate fidelity $F_{\text{gate}}$ is independent of the approximation level $n$ except perhaps for a negligible effect coming from the addition of single qubit gates at the cut locations.  The readout error $R$ may be influenced by $n$ if certain basis have higher associated readout error, however, this effect cannot be predicted without the details of the hardware.   However, there is a clear relationship between the approximation error and the statistical error.  We can see that there is an expected crossover point near $N_C \approx S_{\text{total}}^{3/5}$ for the above example.  Past this crossover point, increasing the approximation level no longer improves the outcome, as it consumes too much of the sample budget.

\end{document}